# Amphiphilic co-polymer network gel films based on tetra-poly(ethylene glycol) and tetra-poly(ε-caprolactone)

**Kevin Hagmann, Carolin Bunk, Frank Böhme and Regine von Klitzing**

+Institute for Condensed Matter Physics, Technische Universität Darmstadt, Hochschulstr. 8, D-64289 Darmstadt
*Leibniz-Institut für Polymerforschung, Dresden e.V, Hohe Str. 6, D-01069 Dresden
#Organic Chemistry of Polymers, Technische Universität Dresden, D-01062 Dresden

**Abstract:** In order to allow the simultaneous transport of hydrophilic and hydrophobic substances, polymeric networks with finely distributed hydrophilic and hydrophobic components are very suitable. When designing new soft materials like coatings, in addition to the structure in the volume phase, the structure at the interface plays a critical role. In this study, two alternating tetra-arm star polymers poly(ε-caprolactone) (tetra-PCL-Ox) and amino-terminated poly(ethylene glycol) (tetra-PEG-NH$_2$) form an amphiphilic co-polymer network. They serve as a model system for controlling and understanding the structure-property relationship at the interface of ACN gels. The correlation between different synthesis strategies for gel films and their resulting properties will be described. Through various spin coating techniques, control over film thickness and roughness is achievable and highlights differences to macroscopic gel samples. Atomic force microscopy (AFM) measurements reveal the effect of solvents of different polarities on the swelling ability and surface structure. This correlates with AFM investigations of the mechanical properties on ACN gel films, demonstrating a strong effect on the resulting elastic modulus E, depending on the presence or absence of a good solvent during synthesis. Furthermore, a higher E modulus is obtained in the presence of the selective solvent water, compared to the non-selective solvent toluene. This observation is explained through selective swelling of the tetra-arm star polymers displaying a different hydrophobicity.

1. Introduction

Amphiphilic co-polymer networks (ACN) consist of both hydrophilic and hydrophobic polymer building blocks, forming a three-dimensional network.[1] They contain finely dispersed hydrophobic and hydrophilic areas within the network. Through the addition of aqueous or organic solvents, the swelling behavior of areas with corresponding hydrophobicity can be selectively manipulated.[1,2] This macroscopic and microscopic change in structure has a significant impact on the transport and diffusion of materials through the network. This behavior is leveraged in soft contact lenses, which is the main industrial application of ACNs.[3–5] Further applications involve membranes for fuel cells, switchable systems for cell cultures, and antimicrobial coatings.[6–8] All of these applications require precise knowledge of the nano- and microscopic structures, as well as the resulting properties in various environments. The mechanical and rheological properties are not only a result of the initial synthesis conditions, but also greatly influenced by partial and selective swelling of ACNs. In order to create tailored polymer networks that exhibit desired mechanical and transport properties, the network structure has to be fully controlled. A model system with a defined mesh structure and network topology is a good starting point to fully understand the structure-swelling-mechanics relationship.

There are viable synthesis strategies to obtain ACNs, however, obtaining networks with defined structures either prove difficult[1] or certain criteria for model networks remain unmet.[9,10] Sakai et al.[11] demonstrated a promising pathway to control polymer network structures and minimize network defects. Through the hetero-complementary coupling of two orthogonal functionalized, four-arm polyethylene glycol stars ($A_4$- and $B_4$-type tetra-PEG), they created a new type of model polymer network. The different end group functionalization prevents one star-type to form bonds with stars of its own type, as well as connections of arms within one star. Ideally, the two star-types bind interchangeably and a network with a defined number of chain segments in all directions is formed. Scattering experiments[12–14], low-field NMR[15], and computer simulations[16] suggest that there are hardly any network inhomogeneities and only a low amount of connectivity errors, resulting in controllable mechanical properties. One shortcoming in these model systems is the presence of some network defects that derive from multiple arm connections between neighboring stars. This results in pending chains and finite loops that can influence mechanical properties.[15]

A better control over the network structure and transfer of tetra-polymer type stars to amphiphilic co-polymer networks has recently been offered by Bunk et al.[42] The network formation is based on the hetero-complementary coupling of hydrophilic and hydrophobic star polymers. The hydrophilic star is composed of an amine end group functionalized tetra-PEG-$NH_2$. The hydrophobic star is composed of oxazinone-terminated poly-$\varepsilon$-caprolactone (tetra-PCL-Ox). Bond formation between two adjacent stars of opposing hydrophobicity is ensured through the reaction of the oxazinone and the amino terminal groups (Figure 1). The advantage of this reaction lies in the great selectivity and control of reaction kinetics, which can be influenced by the variation of concentrations of starting polymers as well as temperature or solvent type. Careful manipulation of these parameters offers great opportunities for a well-defined formation of ACNs.

Great potential for the applications of ACNs can be found in coatings. In addition to the structure in the volume phase, the structure at the interface plays an important role. This is of special interest when dealing with amphiphilic polymers, since their difference in chemical nature can result in a multitude of different structures depending on the environment. Permeability of the material at the interface is one major consideration and depends strongly on the composition of the near-surface region. Methods for investigating surfaces involve atomic force microscopy (AFM), small-angle x-ray scattering, or small-angle neutron scattering under grazing-incidence (GIXAXS, GISANS). These methods allow the investigation of surface topologies that derive from nano- and microphase separations.

The structure of block co-polymers or polymer blends in selective solvents has been investigated with interface-sensitive methods in the past.[17,18] Micro-phase separations at the interface can be visualized by AFM characterization, allowing for the calculation of spectral densities and characteristic lengths. In-situ changes in morphology while changing external parameters were observed using GISAXS.[19] The importance of distinguishing between bulk and surface structure was highlighted by Guzman et al.,[20,21] who demonstrated that gels from co-polymers with a hydrophilic backbone and hydrophobic side chains form a hydrophobic layer at the surface during thin film formation. This hydrophobic layer serves as a barrier for solvent transport in the network. On the contrary, Kamata et al. only observed a homogenous shrinking for ACNs, composed of tetra-PEG and tetra-PEGE, during heating.[22] No indication of changes in surface structure was observed. Other studies show that the polymer-air interface is not the only interface influencing structural changes. Bruns et al. investigated the nanophase separation in films and membranes of ACNs (PDMS and 2-Hydroethylacrlyat), where the surface structure was influenced by the chemical nature of the substrate during network formation.[43] In addition, swelling behavior and type of phase separation highly depend on the composition. Reorganization of the network structure occurred when comparing dry and swollen gels. At this point it is not entirely clear which conditions lead to the formation of a hydrophobic surface layer that varies structurally from the bulk phase. Furthermore, the influence of all barriers in the system and their influence on micro- and nanophase separation is not fully understood.

Preparation conditions for thin films have a potentially great impact on the structure at interfaces. Therefore, synthesis conditions must be controlled precisely. Thin film preparation techniques vary depending on the polymer type. Layer-by-layer methods[23–26], grafting-from-methods[27], or adsorption techniques[28,29] are suitable to prepare stable coatings of polyelectrolyte multilayers, polymer brushes, and microgels, respectively. Other common techniques to obtain thin gel films are dip coating and spin coating. The influence of various parameters during the spin coating process has been studied in detail by Riegler and coworkers.[30–32]

The above-mentioned techniques result in the formation of thin gel films. Similar to the bulk phase, the structure, swelling behavior, and the mechanical and rheological properties are closely related. Atomic force microscopy (AFM) offers the ability to investigate and interpret all these features in close proximity at the interface. A comprehensive overview regarding the determination of elastic properties of materials was provided by Butt et al.[33] The corrections of the Hertz-Model, which is necessary to interpret AFM results, was developed by Dimitriadis et al.[34] Challenges arising from variations in sample properties, such as the thickness of the material, can influence results drastically.[35] Additionally, the geometry of the indenter, which is used to perform elastic measurements, can also have a great impact on the calculations.[33] For homogeneous samples, the use of micrometer-sized colloidal probes is beneficial and easy to model. Samples with heterogeneity require the use of nm-sized AFM-tips that complicate evaluations substantially.[36] However, AFM-tips allow for a high lateral resolution. Surface topology and mechanical properties can be determined in close proximity with resolutions <10 nm, as has been demonstrated in the case of PNIPAM-microgels.[29,37] Differences in elasticity hinted towards a heterogenic gel structure. Furthermore, when the crosslinker content in PNIPAM-microgels is increased, the elastic modulus increases as well.[29] AFM-measurements with high spatial resolution clearly revealed the impact on the mechanical properties of PNIPAM-gels when homogeneous and heterogeneous gels were compared.[38]

With regards to the amphiphilic co-polymer networks investigated in this work, AFM offers a high degree of sample knowledge in terms of surface structure, swelling behavior and mechanical properties. The synthesis, swelling behavior and structure of *bulk gels* of the model amphiphilic co-polymer network, containing interchangeably connected tetra-arm stars tetra- PCL and tetra-PEG, has been recently investigated by Bunk et al.[42] However, the goal of this study is to gain a deep understanding of the structure-swelling-mechanics relationship at the *interface* of these gels. Therefore, the emphasis is to find a suitable preparation technique for thin gel films of ACNs. For comparison, a macroscopic gel film (thickness ~ mm) serves as a reference interface, whose structure should not vary from the interface of the volume phase. Control of surface morphology, roughness, film thickness, and customized sample stiffness and elasticity is highly desirable. Atomic force microscopy (AFM) with its high resolution serves as the main tool for surface analysis to evaluate the influence of various synthesis protocols or different chemical environments gel films are subjected to.

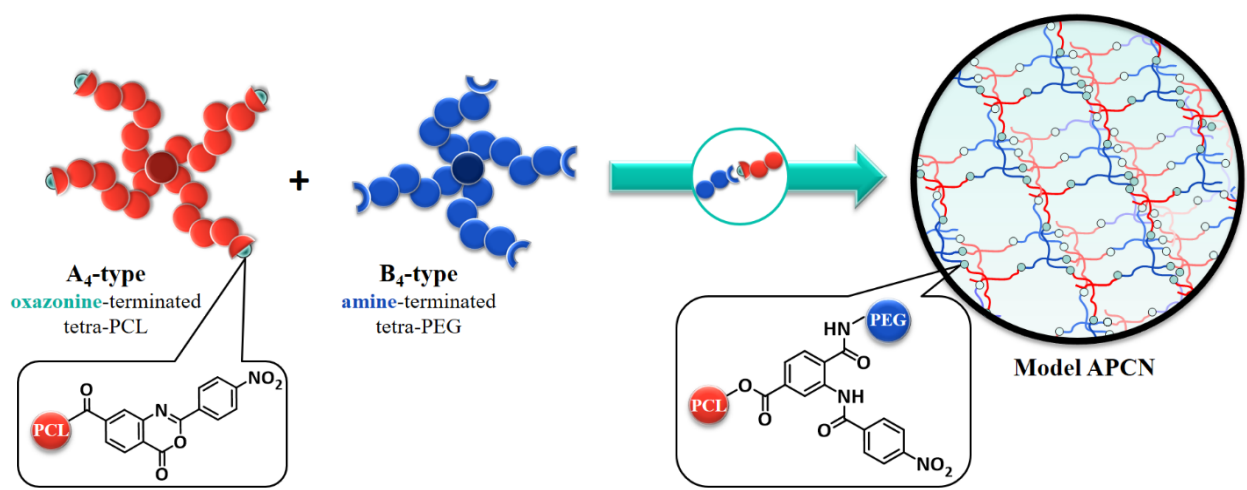

**Figure 1.** Synthesis of ACNs from hydrophobic 2-(4-nitrophenyl)-benzoxazinone-terminated PCL and hydrophilic amine-terminated PEG tetra-arm star polymers.[42]

## 2. Materials and Methods

*2.1 Materials*

Sulfuric acid (100%) and hydrogen peroxide (30%) were purchased from Carl Roth (Germany). 3-amino-propyltriethoxysilane (99%) (APTES) was purchased from Sigma Aldrich (Germany). Toluene (99.7%) was purchased from Honeywell (Germany). Milli-Q water with a resistance of 18 MΩcm was produced in an in-house Milli-Q-system from Merck.

*2.2 Synthesis of amine-terminated tetra-PEG (tetra-PEG-NH$_2$) and 2-(4-nitrophenyl) benzoxazinone-terminated tetra-PCL (tetra-PCL-Ox)*

Both, amine-terminated star-shaped poly(ethylene glycol) (MW = 10 kg mol$^{-1}$) and 2-(4-nitrophenyl)-benzoxazinone-terminated tetra-arm star-shaped poly(ε-caprolactone) (MW = 10 kg mol$^{-1}$) were prepared by procedures previously established by Bunk et al.[42]

*2.3 Preparation of substrates*

*2.3.1 Silicon or glass substrates*

Silicon wafers from Siegert Wafer (Aachen, Germany) and glass slides from LABSOLUTE (Geyer GmbH, Renningen, Germany) were cleaned thoroughly in acetone, isopropanol, and Milli-Q water and subsequently etched in piranha solution (3:1 H$_2$SO:H$_2$O$_2$) for 20 mins. After rinsing with Milli-Q water, substrates were dried in a stream of nitrogen gas.

*2.3.2 Amino-functionalized substrates*

Silicon wafers from Siegert Wafer (Aachen, Germany) or glass substrates were silanized with 3-amino-propyltriethoxysilane (APTES) similarly to established procedures.[39,40] In short, substrates were cleaned thoroughly in acetone, isopropanol, and Milli-Q water and subsequently etched in piranha solution (3:1 H$_2$SO:H$_2$O$_2$) for 20 mins. After rinsing with Milli-Q water, substrates were dried in a stream of nitrogen gas. The clean substrates were immersed into APTES-toluene solution (10 v%) and kept for 2h at room temperature. Afterwards, substrates were washed with toluene to remove excess APTES and dried in a stream of nitrogen gas. Substrates were pre-heated to 60 °C in an oven and immediately used for the following coating procedures.

*2.4 Preparation of ACN gel films*

All ACN gel samples were prepared from stock solutions of the tetra-arm star polymers tetra-PCL-Ox and tetra-PEG-NH$_2$, respectively. The polymers were dissolved individually at a concentration of 70 mg/mL in toluene at 60 °C for 30 min. The ACN reaction was initiated by mixing both solutions in equimolar ratio (for details see Ref. 42). The solution was stirred for 20 min at 60°C and at that point further processed into thin films or thicker samples before gelation could occur (after around 30 min).

*2.4.1 Preparation of thin ACN gel films*

Thin films of ACNs were prepared by spin-coating with a spin coater from SETCAS LLC, California, USA. After mixing equimolar aliquots of the tetra-arm stars stock solutions and a reaction time of 20 min at 60 °C, 200 µL of the reaction mixture was spin-coated onto chemically modified silicon wafers (preheated to 60°C). Spin coating procedures were carried out at rotation speeds of 500 – 7000 rpm for 1 min, either through static or dynamic spin coating. For *static spin coating*, the reaction mixture was pipetted onto a silicon wafer before initiation of rotation. For *dynamic spin coating*, the reaction mixture was pipetted quickly onto a silicon wafer that had already been accelerated to the desired rotation speed. The coated wafers were afterwards kept in an oven at 60 °C overnight to finalize the reaction. Samples were dried in a vacuum oven before analysis.

*2.4.2 Preparation of bulk ACN gel films*

After the initial ACN mixture has reacted at 60 °C for 20 min, the solution was poured into a mold onto a chemically modified glass or silicon substrate and sealed in an airtight Teflon vessel to prevent further evaporation. The samples were kept in an oven at 60 °C overnight to finalize the reaction. Samples were either analyzed directly in toluene or dried before analysis.

*2.5 AFM*

ACN gel films were characterized with atomic force microscopy (AFM). AFM measurements were carried out at ambient conditions on either the MFP 3D SA or Cypher (both, Asylum Research/Oxford Instruments, Wiesbaden, Germany). First, AFM was used in tapping mode to obtain information about surface topographies of ACN gel films. The cantilevers AC160TS-R3 (300 kHz, 26 N m$^{-1}$) were used for scanning dried samples in ambient conditions, while the cantilevers AC240TS-R3 (70 kHz, 2 N m$^{-1}$) were used for scanning samples in water or toluene. Both were purchased from Asylum Research/Oxford Instruments, Wiesbaden, Germany. Secondly, AFM was used in static force measurements. Cantilevers were chosen to best match the elasticity of the sample. The cantilevers NSC18 (75 kHz, 2.8 N m$^{-1}$) were used for samples exhibiting elastic moduli in the low MPa region, the cantilevers CSC37/No Al (20 kHz, 0.3 N m$^{-1}$) for elastic moduli between 100 – 1000 kPa, and CSC38/No Al (10 kHz, 0.03 N m$^{-1}$) for elastic moduli smaller than 100 kPa. All cantilevers for indentation measurements were fabricated by MikroMasch and purchased from NanoAndMore, Wetzlar, Germany.

Film roughness and film thickness were determined through analysis of surface topographies. The roughness of a recorded 20 x 20 μm$^2$ image was determined using the built-in software features of IGOR 16 (Asylum research). Images were recorded for all preparation conditions with at least threefold repetition. A 20 x 20 μm$^2$ image close to the center of each sample was chosen for evaluation. Randomly chosen locations with a box size of 1 x 1 μm$^2$ on any recorded image were evaluated and averaged against results from repeated experiments to obtain roughness values and standard deviations. The thickness of a film was determined by scratching the sample close to the center of the substrate with a sharp scalpel and scanning at least two different 50 x 50 or 70 x 70 μm$^2$ areas along the cut. The silicon wafer surface and the polymer gel surface were easily distinguishable and the height differences of both surfaces yielded the film thickness. At least three different randomly chosen locations on each recorded image were chosen to obtain film thicknesses. Experiments were repeated at least threefold to obtain the average film thickness and standard deviation.

Elastic moduli were obtained through static force measurements by recording force maps (100 measurements across a 20 x 20 μm$^2$ area) at a minimum of two sample locations on any given sample. Force curves were fitted using the Hertz model[41], using the AFM in-built software features of IGOR 16 (Asylum research). All elastic moduli determinations are based on a Poisson-ratio of 0.5. Samples are obtained from at least threefold repetition of each experiment. Force maps recorded on each sample are used to obtain a distribution of elastic moduli, as well as the average elastic moduli and standard deviations.

## 3. Results

*3.1 Challenges of ACN gel film preparation.*

The bulk reaction, followed up by pouring the reaction mixture into a mold, resulted in the formation of a thick gel film. Gel films were dried under vacuum to investigate their structural changes and swelling behavior after the addition of solvents of different polarity. Upon addition of the selective solvent water or the non-selective solvent toluene to a dried gel film, the gel networks swell and deform macroscopically. Images and a video of a bulk gel film and its deformation during the addition of the selective solvent water can be found in the *supporting information (S)*. After deposition of a water droplet, the water droplet sits on top of the gel for a few seconds, then slowly penetrates and swells the gel and ultimately deforms the gel entirely (Figure S1). Thin films also demonstrated deformation and detachment from the substrate after solvent addition (Figure S2). In contrast, thin and thick gel films that were prepared and immobilized on amino-functionalized (by silanization with APTES) substrates did not deform when water or toluene were added. All subsequent measurements were performed on immobilized gel films.

*3.2 Surface topography and thickness of ACN gel films in ambient conditions*

AFM measurements in tapping mode were used to determine the surface topography, surface roughness, and film thickness of thin gel films on amino-functionalized silicon substrates. Figure 2 shows the surface topography of thin gel films in ambient conditions prepared via static spin coating (2a) and dynamic spin coating (2b), for spin coating speeds of 1000 and 5000 rpm, respectively. Both, height and phase images are displayed. A bulk gel film serves as comparison (2c).

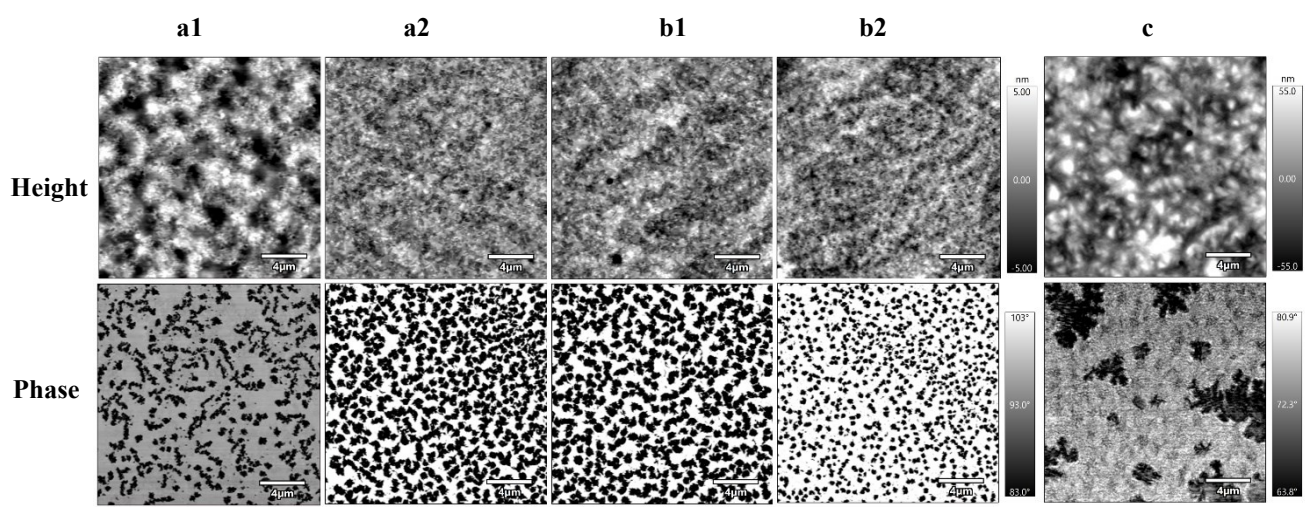

**Figure 2.** Surface topography of ACN gel films obtained by AFM. **a**) Thin film obtained through static spin coating, **b**) Thin film obtained through dynamic spin coating, **c**) bulk gel film. **1**) spin coating speed of 1000 rpm, **2**) spin coating speed of 5000 rpm. The top row displays the AFM height images, the bottom row the AFM phase images

The surface topography obtained from height imaging in AFM tapping mode does not reveal any major differences between prepared thin films, except for the sample prepared at 1000 rpm through static spin coating (2a1). Both spin coating techniques provide similar film surfaces. The surface appears to be overall homogenous with no identifiable super-structures. Phase images are analyzed to gain additional information on the surface composition. Again, there is no clear distinction between different coating techniques. However, the phase images of thin films reveal the presence of two clearly distinguishable structural domains. Phase images can provide qualitative information on the sample-tip interaction, that can be attractive or repulsive, resulting in a difference in phase shift of the cantilever oscillation. Since the ACN gels in this study have a hydrophobic and hydrophilic component, a different interaction between these components and the cantilever tip is to be expected. This makes phase imaging a useful tool to identify different domains of tetra-PEG and tetra-PCL rich or poor areas, respectively. The synthesis of bulk gel films serves as a comparison to establish the connection between bulk and surface structure. The surface topography of the bulk gel film appeared similar to thin films obtained from static spin coating at lower speeds (compare 2a1 and 2c), as determined by the height image, although with an overall higher surface roughness. The average surface roughness of bulk gel films under ambient conditions was 17 ± 5 nm. In contrast to thin films, the phase images of bulk gel films revealed the presence of larger structures, which could potentially indicate polymer aggregates.

Thin films were prepared at rotation speeds of 500 – 7000 rpm by static and dynamic spin coating (Figure 3). Thin films obtained via static spin coating exhibit an average surface roughness of 7 ± 2 nm, which is mostly independent of the rotation speed. Films prepared via dynamic spin coating exhibit an average surface roughness of 5 ± 2 nm, which is again mostly independent of the rotation speed. Static and dynamic spin coating provide a similar film roughness across the board. Cutting into thin films with a scalpel and scanning across the cutting edge allowed for the determination of film thicknesses. Static spin coating produced films with a thickness of roughly 800 – 900 nm depending on the rotation speed. Dynamic spin coating produced a wider range of film thicknesses from 300 – 900 nm, with higher rotation speeds noticeably decreasing film thickness. Dynamic spin coating at lower rotation speeds (≤ 2000 rpm) yielded very inhomogeneous films. The area of impact of the drop addition during the dynamic spin coating process was clearly distinguishable from the rest of the sample with optical microscopy. This resulted in a wide range of film thicknesses, depending on the measured sample spot location. The effects of spin coating technique and rotation speed on film thickness and roughness is summarized in Table 1. The comparison bulk gel films had a thickness of approximately 1 mm. Film thicknesses of bulk films were not determined more precisely by the scratching method, since the large created edge exceeds the scanning capabilities of the AFM instrumentation by far.

**Table 1.** Effects of spin coating technique and rotation speed on film thickness and roughness

| Rotation speed (rpm) | Film thickness (nm) | | Film roughness (nm) | |
|---|---|---|---|---|
| | Static | Dynamic | Static | Dynamic |
| 500 | 900 ± 100 | 800 ± 200 | 6 ± 2 | 6 ± 4 |
| 1000 | 890 ± 60 | 700 ± 300 | 7 ± 4 | 4 ± 1 |
| 2000 | 900 ± 100 | 500 ± 200 | 6 ± 5 | 3 ± 1 |
| 5000 | 880 ± 30 | 470 ± 40 | 6 ± 3 | 6 ± 2 |
| 7000 | 810 ± 30 | 360 ± 40 | 9 ± 4 | 7 ± 3 |

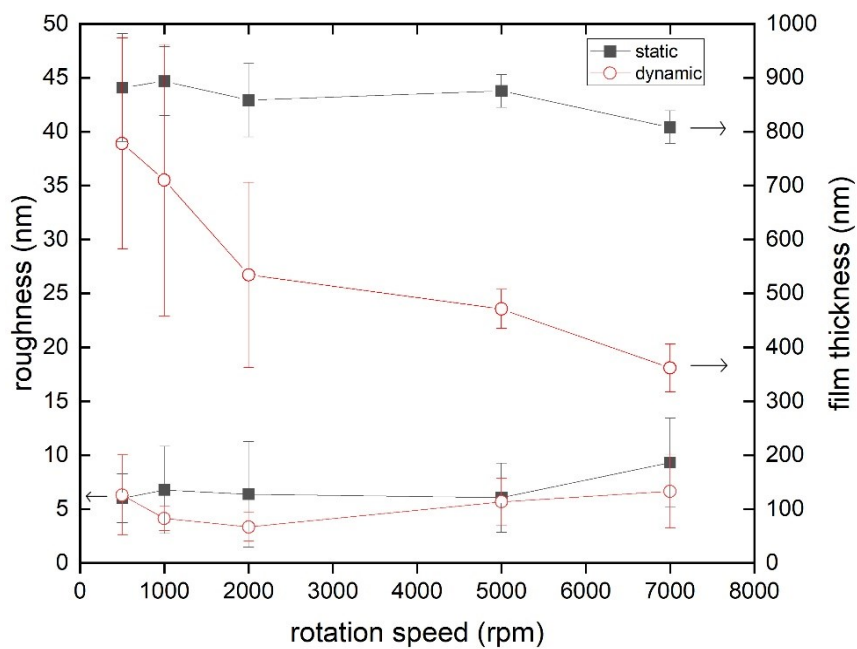

**Figure 3.** Film roughness (bottom two graphs) and thickness (top two graphs) of ACN gel films as a function of rotation speed. Thin films were prepared by static (squares) or dynamic (circles) spin coating

*3.3 Surface topography and thickness of ACN gel films against solvents*

The surface topography changes when the chemical environment of the ACN gel films is varied (Figure 4). At first, water as a selective solvent is added to dried gel films (4b). The surface topography at the gel-water interface for thin films reveals an increase in average surface roughness from 7 ± 4 nm to 35 ± 21 nm for thin films and from 17 ± 5 nm to 76 ± 27 nm for bulk gel films, respectively. The films also swell. For thin films, this is indicated by an increase of film thickness on average by a factor of 1.5, compared to dry gel samples. Film thicknesses for bulk gel samples could not be determined due to the AFM instrument limitations described above. The general structure of the gel-water interface (4b) does not vary significantly from the gel-air interface (4a), except for the increase in roughness. This observation holds true for both thin films and bulk gel films.

Secondly, the non-selective solvent toluene is added to dried gel films (4c). The surface topography at the gel-toluene interface reveals a significant increase in surface roughness from 7 ± 4 nm to 35 ± 17 nm for thin films and from 17 ± 5 nm to 66 ± 27 nm for bulk films, respectively. For thin films, the swelling ability is demonstrated by an increase of film thickness by a factor of 2.4. ACN thin films show a greater degree of swelling in toluene than in water. Noteworthy is the change of surface structure in toluene compared to ambient conditions. Beyond the increase in roughness, grooves with depths of up to 300 nm appear along the entire gel surface.

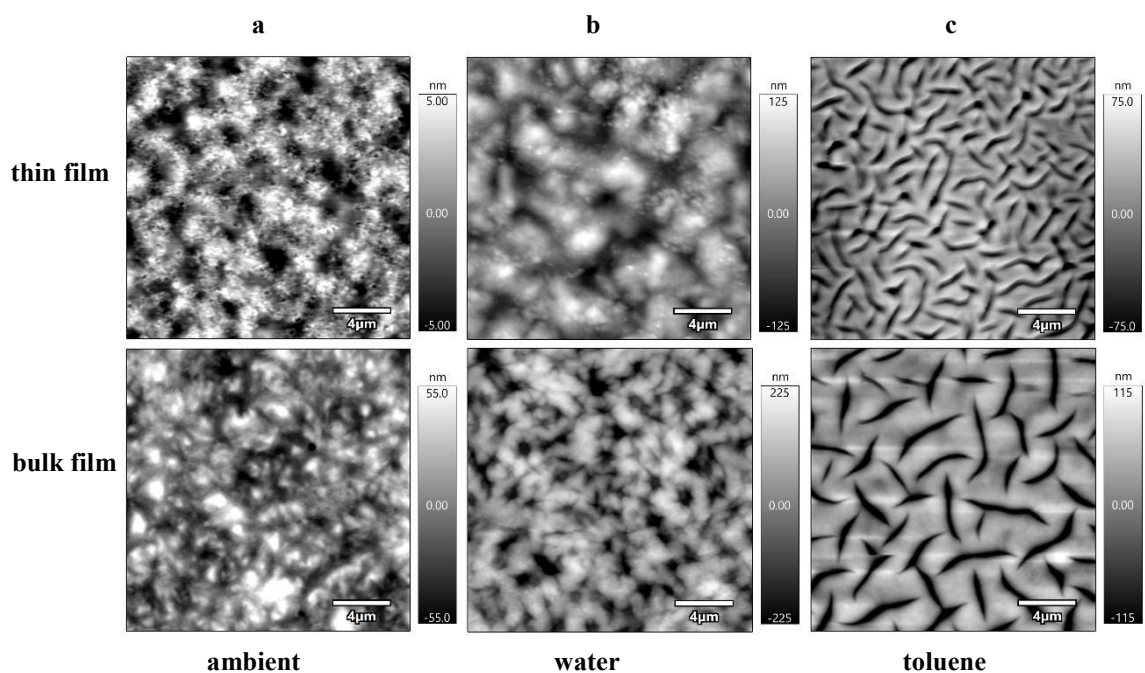

**Figure 4.** Surface topography of ACN gel films obtained by AFM (height image) under the respective condition. **a)** Gel film under ambient conditions, **b**) Gel film swollen in water, **c**) Gel film swollen in toluene. The top row displays images obtained from a thin film (static spin coating, 1000 rpm). The bottom row displays images obtained from a bulk gel film

**Table 2.** Surface roughness of thin and bulk gel films in ambient conditions, water, and toluene

| Sample | Film roughness (nm) Ambient conditions | Film roughness (nm) Water | Film roughness (nm) Toluene |
| --- | --- | --- | --- |
| Thin film | 7 ± 4 | 35 ± 21 | 35 ± 17 |
| Bulk film | 17 ± 5 | 76 ± 27 | 66 ± 27 |

*3.4 Mechanical and properties of ACN gel films*

AFM indentation measurements with a tip (tip radius < 30 nm) were used to determine the elastic modulus of gel films at the surface in the presence of the selective solvent water or non-selective solvent toluene. The average elastic moduli of thin gel films (tf), irrespective of preparation technique, are $E_{tf,w}$ = 2800 ± 200 kPa in water (w) and $E_{tf,t}$ = 400 ± 100 kPa in toluene (t). The average elastic moduli of bulk gel films (bf) are $E_{bf,w}$ = 290 ± 80 kPa in water and $E_{bf,t}$ = 93 ± 6 kPa in toluene. A summary is shown in Table 3. For all sample types, elastic moduli are displayed as a distribution (Figure 5), which is necessary due to the small tip radius. As expected, the elastic moduli of gel films in the selective solvent water are higher than in the non-selective solvent toluene. The elastic moduli of water samples also display a broader distribution of the elastic modulus compared to toluene samples. Bulk gel films exhibit a significantly lower elastic modulus compared to the thin gel films.

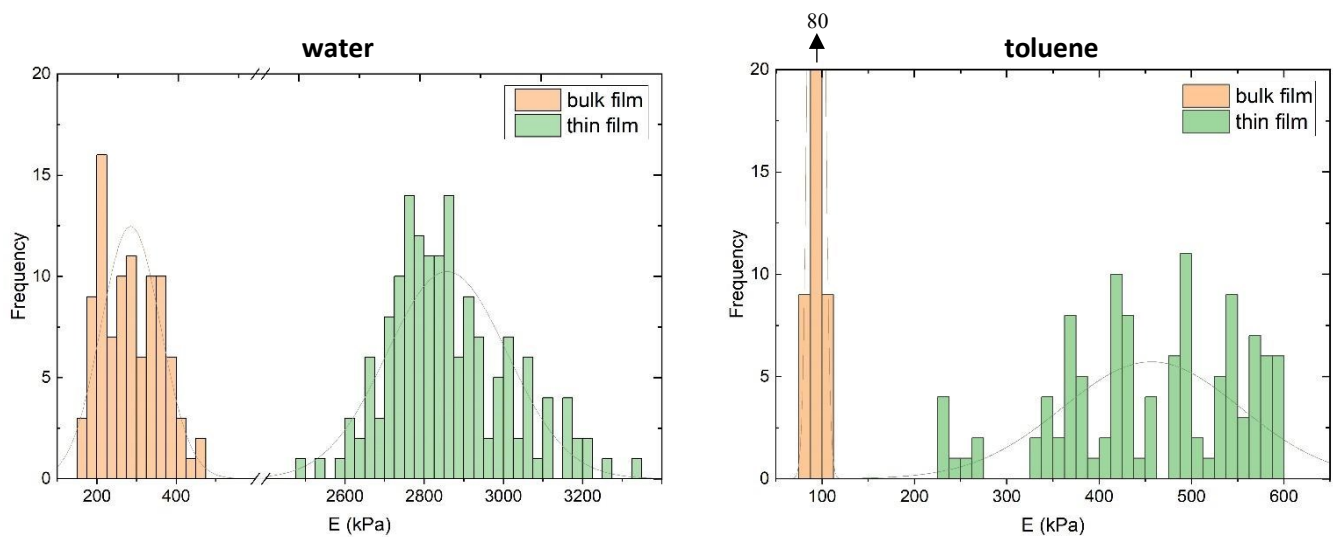

**Figure 5.** Distribution of elastic moduli of ACN thin films (green) and bulk films (orange) in the selective solvent water (left) and the non-selective solvent toluene (right), obtained by AFM indentation experiments with tip (tip radius < 30 nm)

**Table 3.** Elastic moduli E of ACN thin films and bulk films in the selective solvent water and the non-selective solvent toluene, obtained by AFM indentation experiments with tip (tip radius < 30 nm)

| Sample | $E_{water}$ (kPa) | $E_{toluene}$ (kPa) |
|---|---|---|
| Thin film | $2800 \pm 200$ | $400 \pm 100$ |
| Bulk film | $290 \pm 80$ | $93 \pm 6$ |

## 4. Discussion

Firstly, the preparation technique of gel films is discussed. Next the structure of the resulting ACN gel films at the surface is evaluated. Lastly, the mechanical and rheological in different chemical environments is analyzed.

### 4.1 Preparation of ACN gel films

The synthesis of ACN gel films consisting of tetra-PEG-NH$_2$ and tetra-PCL-Ox is highly tunable, since the reaction is extremely sensitive to concentration, temperature, and solvent type. This is a consequence of the different solubilities of the hydrophilic and hydrophobic polymer starters, as well as the variable reaction kinetics of the oxazinone-amine reaction used.[42] Therefore, any deviation from established synthesis conditions has a potentially great impact on the structure of the gel film, as demonstrated by differences in one-arm or multiple-arm connections via low field NMR – measurements.[42] The approach chosen in this study allows the synthesis of thin gel films that should mimic the bulk reaction as closely as possible. The non-selective solvent toluene ensures similar solubility of both polymer starters and a temperature of 60 °C avoided all potential demixing or crystallization events, once the solvent is removed. For the thin gel formation, ACNs were pre-formed in a bulk reaction over a time period until shortly before the gelation of the ACN[42] to ensure similarity to the structure of bulk gels. Before gelation occurs, samples needed to be spin-coated to obtain smooth thin films that allow access to surface-sensitive analyses. After finishing the reaction in an oven, thin gel films were obtained.

However, when adding the selective solvent water or non-selective solvent toluene, samples deform drastically and detach from the surface, making characterization by AFM measurements impossible. Therefore, it is necessary to immobilize gel films onto the substrate. The use of APTES leads to the amino-functionalization of silicon or glass substrates, and subsequently allowed the formation of covalent linkage of the ACN gel to the substrate. The underlying chemical reaction is equal to the linking procedure of the different star polymers. The coupling agent attached to the arms of tetra-PCL can bind to the APTES-$NH_2$ -groups on the substrate surface. Gel films were successfully immobilized and did not deform after the addition of solvents.

*4.2 AFM topography of ACN gel surfaces*

The surface topography of thin ACN gel films in ambient conditions, obtained by AFM measurements, demonstrates the influence of dynamic and static spin coating. Both techniques lead to similar surface structures, and the measured surface roughness is matching closely. However, when trying to control the thickness of the resulting gel film, dynamic spin coating is a superior method in this case. Film thicknesses for dynamic spin coating almost linearly decrease with increase of rotation speed. In our system, dynamic spin coating should only be applied with rotation speeds of 5000 rpm and higher. At lower rotation speeds, film thicknesses vary a lot depending on spot location on a given sample, as demonstrated by the high standard deviation for lower rotation speeds. Overall, gels with a film thickness of 300 - 900 nm can be prepared. The thickness of the APTES layer is neglected due to it being less than 3 nm.[39] Both methods fall short in creating thin films sub 200 nm, which some surface sensitive measurements require (e.g. x-ray or neutron reflectometry). In comparison, the bulk gel film has a similar surface structure, but a much higher surface roughness. This again highlights the importance of thin film preparation techniques, since rough samples complicate surface sensitive measurements. The phase images in ambient conditions reveal the presence of larger domains of inhomogeneities. This can be combined with the observation that water droplets sit on the dry gel surface before they slowly penetrate the gel and the water droplet spreads. One could argue the presence of hydrophobic domains at the gel-air interface, that prevent transport of solvents, as has been reported by Guzman et al.[20,21] In contrast, toluene immediately spreads and penetrates gel films.

The swelling behavior and impact on the surface structure are investigated by the addition of water or toluene. On one hand, the addition of water leads to an increase of surface roughness and an increase of film thickness. These observations are explained by the selective swelling of tetra-PEG stars in the gel network, increasing both roughness and film thickness. On the other hand, the use of toluene results in a similar increase in surface roughness but a higher film thickness compared to water. Since toluene is a good solvent for tetra-PEG and tetra-PCL in the network, both components are subject to swelling, explaining the higher film thickness. In addition, toluene changed the surface structure significantly for both thin films and bulk films, displaying rather compact domains separated by deep grooves across the entire gel surface.

*4.3 Nano-mechanical properties at ACN gel surfaces*

AFM indentation experiments reveal significant differences in mechanical properties depending on sample and solvent type. Generally, ACN gel films in toluene demonstrated elastic moduli an order of magnitude lower than gel films in water. This observation qualitatively coincides with postulated and measured swelling behavior of ACN gel films. As discussed previously, toluene is a good solvent for both star types. Film thicknesses of thin gel films increase by a greater amount compared to the selective solvent water due to swelling. Film thicknesses of dried thin gel films increase by a factor of 2.4 in toluene and by 1.5 in water, respectively. A more swollen gel would therefore result in a lower elastic modulus, which is confirmed by our experimental results. Additionally, elastic moduli measured for toluene samples show a much narrower distribution compared to water samples. This is expected since equal swelling of both star types in toluene would potentially result in a more homogeneous structure of gels. Meanwhile, the selective solvent water only partially swells the hydrophilic tetra-PEG stars, leaving the tetra-PCL stars in a collapsed state, potentially aggregating and forming super-structures.

Another noteworthy observation is the significant difference of elastic moduli measured in thin gel films compared to the bulk gel film, again by an order of magnitude. However, the surface topography of the corresponding gels and swelling behavior do not immediately suggest any structural differences between thin gel films and bulk gels. Substrate influences can be excluded, since all thin films independently of thickness have comparable elastic moduli. This magnitude of variation in mechanical properties must therefore derive from the initial synthesis conditions. The bulk gel film is synthesized entirely in the initial bulk reaction and even after pouring into a mold maintains its concentration of tetra-PEG and tetra-PCL stars, respectively. In contrast, for the thin film preparation, the solvent is almost entirely removed by spin coating after the initial 20 min reaction in the bulk. At this point, less than 50% of end groups have reacted, according to NMR results.[42] This means that the remaining reaction takes place in a state of a melt with a much higher local concentration and lower diffusion rates when unreacted end groups try to find a matching partner. But most importantly, different levels of the network can now interconnect, and therefore greatly reduce the overall size of the network and increase its inhomogeneities. This results in more tension in the network as well as a reduced swelling ability. When determining elastic properties, a higher elastic modulus would therefore be expected.

## 5. Conclusion

This study concentrates on the development of a suitable preparation technique of thin ACN gel films, consisting of a model system with four-arm tetra-PEG and tetra-PCL stars linked by an oxazinone-amine coupling reaction. The focus lies on the analysis of the structure at the surface or more general interfaces of gel films and how it compares to the bulk structure. This model system should serve as a starting point to gain control and understanding of the structure-mechanics-rheology relationship at the interface. Ultimately, this would allow the design of soft gels where properties such as swelling behavior, permeability, and transport properties are tunable.

Gel films are prepared via bulk reaction followed by dynamic or static spin coating or pouring into a mold. All gel films exhibit deformation and detachment from the substrate during the addition of the selective solvent water and the non-selective solvent toluene. Preventing these phenomena is achieved by chemical immobilization onto the silica or glass substrate with APTES, therefore allowing the use of surface sensitive characterization methods. Surface topography, roughness, and film thickness of ACN gel films are characterized by atomic force microscopy. Both static and dynamic spin coating result in the formation of gel films with similar surface structure in ambient conditions with low surface roughness. Dynamic spin coating is the superior method for film thickness control, but may result in less homogeneous films depending on the rotation speed. All samples show an increase in roughness and film thickness during the addition of water or toluene. Samples in the non-selective solvent toluene had higher are thicker compared to samples in the selective solvent water. This is a result of the different hydrophobicities and swelling capabilities of tetra-PEG (hydrophilic) and tetra-PCL (hydrophobic). The surface topographies reveal the presence of different structure types, potentially being a result of the aggregation of same star species. Gel films in toluene display the formation of compact surface domains separated by grooves. Interestingly, thin gel films and bulk gel films behave similarly up to this point, potentially indicating similar chemical structures.

AFM was further used to evaluate the nano-mechanical and nano-rheological properties of ACN gel films in water and toluene. The elastic modulus of samples in water is significantly higher than of samples in toluene, again confirming the higher degree of swelling of an amphiphilic network in a non-selective solvent, compared to a selective solvent. Static AFM indentation measurements revealed great differences between thin gel films and bulk films. We propose the differences arising from the preparation method. The early removal of solvent by spin coating to prepare thin films results in a modified state that the reaction proceeds in. Molten state, lower diffusion rates, higher network tension, and interconnection of different network levels result in a much stiffer gel network. This could be partially remedied by the reswelling in solvents, but not completely avoided. We are faced with a dilemma: On one hand, we are dealing with a highly tunable system, where the initial concentration, temperature and type of solvent are adjustable parameters. This would potentially allow the formation of "perfect" networks. However, on the other hand, this highly sensitive behavior results in challenges when trying to establish comparable sample types both on a macroscopic and microscopic level. The availability of smooth thin films is of utmost importance for surface-sensitive analytical methods. Further adjustment of preparation parameters might help to tailor the structure, swelling behavior and mechanical properties of the ACN films.

**Acknowledgement:**

This study was conducted as part of the research collaboration "Adaptive Polymer Gels with Model-Network Structure" (FOR2811), funded by the German Research Foundation (DFG), grant 423768931, 423514254 and 397384169.

The authors thank Hans Riegler for discussions about different spin coating procedures.

# Supporting Information

Amphiphilic co-polymer network gel films based on tetra-poly(ethylene glycol) and tetra-poly(ε-caprolactone)

Kevin Hagmann[+], Carolin Bunk[*,#], Frank Böhme[*] and Regine von Klitzing[+]

[+]Institute for Condensed Matter Physics, Technische Universität Darmstadt, Hochschulstr. 8, D-64289 Darmstadt
[*]Leibniz-Institut für Polymerforschung, Dresden e.V, Hohe Str. 6, D-01069 Dresden
[#]Organic Chemistry of Polymers, Technische Universität Dresden, D-01062 Dresden

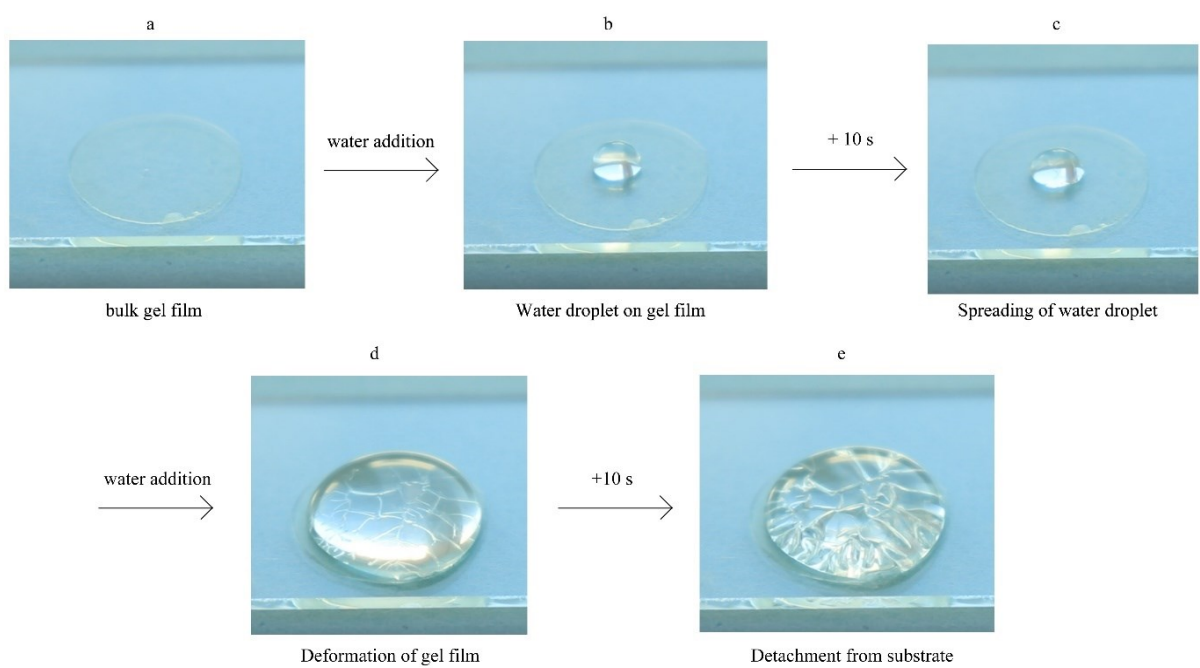

Figure S1: Swelling behavior of gel film in water. a) Dried bulk gel film, b) Pipetting of water droplet onto gel film, c) After a few seconds the water droplet noticeably penetrates the network, the water droplet spreads/contact angle decreases. d) The swelling and deformation of the gel film is better visualized through the addition of additional water. e) The gel film fully deforms and detaches from the glass substrate

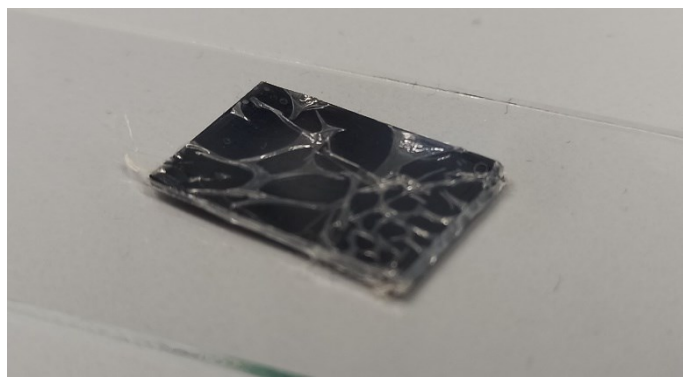

Figure S2: Deformation of ACN thin gel film (thickness ~ µm) and detachment from silicon substrate after the addition of water.